# Eigenspace-Based Minimum Variance Adaptive Beamformer Combined with Delay Multiply and Sum: Experimental Study

Moein Mozaffarzadeh[a], Ali Mahloojifar[a], Mohammadreza Nasiriavanaki[b], and Mahdi Orooji[a]

[a]Department of Biomedical Engineering, Tarbiat Modares University, Tehran, Iran
[b]Wayne State University, Bioengineering Department, Detroit, Michigan, United States

**ABSTRACT**

Delay and sum (DAS) is the most common beamforming algorithm in linear-array photoacoustic imaging (PAI) as a result of its simple implementation. However, it leads to a low resolution and high sidelobes. Delay multiply and sum (DMAS) was used to address the incapabilities of DAS, providing a higher image quality. However, the resolution improvement is not well enough compared to eigenspace-based minimum variance (EIBMV). In this paper, the EIBMV beamformer has been combined with DMAS algebra, called EIBMV-DMAS, using the expansion of DMAS algorithm. The proposed method is used as the reconstruction algorithm in linear-array PAI. EIBMV-DMAS is experimentally evaluated where the quantitative and qualitative results show that it outperforms DAS, DMAS and EIBMV. The proposed method degrades the sidelobes for about 365 %, 221 % and 40 %, compared to DAS, DMAS and EIBMV, respectively. Moreover, EIBMV-DMAS improves the SNR about 158 %, 63 % and 20 %, respectively.

## 1. INTRODUCTION

Photoacoustic imaging (PAI), unlike the X-ray which uses an ionizing radiation, is a nonionizing medical imaging modality, providing the resolution of Ultrasound (US) imaging and the contrast of optical imaging.[1] Photoacosutic (PA) signals are generated as a result of a laser illumination toward the target where PA waves are propagated based on the thermoelastic effect and thermal gradient.[2] PAI is a multiscale imaging modality that has been used in different fields of study such as tumor detection[3] and functional imaging.[4]

Since there is a high similarity between US and PA detected signals, many of beamforming algorithms used in US imaging can be used in PAI.[5,6] Moreover, integrating these two imaging modalities has been a challenge over the past few years.[7,8] The most prevalent beamformer in PAI is delay and sum (DAS). While it leads to a low quality image as a result of its nonadaptiveness and blindness, it has been used extensively in PAI due to its simple implementation.[9] To address the incapabilities of DAS, adaptive beamformers, such as minimum variance (MV), are commonly employed. MV has been investigated over the past few years, especially in US imaging.[10–12] Delay and standard deviation (DASD) beamforming algorithm was proposed to address the poor appearance of interventional devices such as guide wires, needles, and catheters in US images.[13]

Matrone *et al.* proposed a new beamforming algorithm, called delay multiply and sum (DMAS), as a beamforming technique for linear array US imaging.[14] Initially, DMAS was used for breast cancer detection, introduced by Lim *et al.*[15] DMAS has been used with multi-line transmission (MLT) for high frame-rate US imaging.[16] Double stage (DS-DMAS), in which two stages of DMAS is used in order to achieve higher contrast and resolution compared to DMAS, was proposed based on the expansion of DMAS.[17,18] The lack of a high resolution in DMAS was addressed using combining MV with DMAS, and it was shown that in the expansion of DMAS, there are some terms which can be interpreted as DAS. These terms can be replaced by MV in order to improve the resolution while the contrast is retained.[19–21] A modified version of coherence factor (MCF) was introduced

---

based on DMAS, and it was shown that it provides higher contrast compared to conventional CF.[22]

In this paper, a novel algorithm based on combination of DMAS and eigenspace-based MV(EIBMV) is proposed. The algorithm is called EIBMV-DMAS, and we have numerically evaluated this algorithm before, and the results can be found in reference.[20] The expansion of DMAS algorithm was used, and it was shown that in each term of the expansion, there is a DAS algebra. Since DAS algorithm is a non-adaptive beamformer and leads to low resolution images, we proposed to use EIBMV instead of the existing DAS in the expansion of DMAS. Here, we evaluate the proposed method experimentally. The experimental results show that EIBMV-DMAS provides resolution improvement and levels of sidelobe reduction, compared to DMAS and EIBMV, at the expense of higher computational burden.

The rest of the paper is as follows. In section 2, the concept of beamforming in linear-array imaging and the proposed method are presented. Section 3 contains the experimental results and the necessary evaluation. Finally, the conclusion is presented in section 4.

## 2. MATERIALS AND METHODS

In this section, the concept of the proposed algorithm is explained. When PA signals are detected by a linear array of US transducer, beamforming algorithms such as DAS can be used to reconstruct the image from the detected PA signals. Formula of DAS, as the most common beamformer, can be written as follows:

$$y_{DAS}(k) = \sum_{i=1}^{M} x_i(k - \Delta_i), \tag{1}$$

where $y_{DAS}(k)$ is the output of the beamformer, $k$ is the time index, $M$ is the number of elements of array, and $x_i(k)$ and $\Delta_i$ are the detected signals and the corresponding time delay for detector $i$, respectively. DAS results in low quality images due to its blindness and also the limited available angle view in linear-array imaging. To address this problem, DMAS was introduced in which the same as DAS, corresponding sample for each element is calculated, but samples are combinatorially coupled and multiplied before summation.[14] DMAS formula is given as follows:

$$y_{DMAS}(k) = \sum_{i=1}^{M-1} \sum_{j=i+1}^{M} x_i(k - \Delta_i) x_j(k - \Delta_j). \tag{2}$$

The dimensionally squared problem of (2) was also addressed using following equations:[14]

$$\hat{x}_{ij}(k) = \text{sign}[x_i(k-\Delta_i)x_j(k-\Delta_j)]\sqrt{|x_i(k-\Delta_i)x_j(k-\Delta_j)|}, \quad \text{for} \quad 1 \leqslant i \leqslant j \leqslant M. \tag{3}$$

$$y_{DMAS}(k) = \sum_{i=1}^{M-1} \sum_{j=i+1}^{M} \hat{x}_{ij}(k). \tag{4}$$

Performing sign, absolute and square root after coupling procedure in (3) and (4), which requires $(M^2 - M)/2$ computations for each pixel, result in a slow imaging. To put it more simply, sometimes these library functions require many clock cycles, causing improper timing performance of DMAS algorithm. Applying following procedure to the received PA signals reduces the computational number of the sign, absolute and square root operations to $M$ for each pixel:[23]

$$\bar{x}_i(k) = \text{sign}[x_i(k)]\sqrt{x_i(k)} \quad \text{for} \quad 1 \leqslant i \leqslant M, \tag{5}$$

$$\hat{x}_{ij}(k) = \bar{x}_i(k)\bar{x}_j(k) \quad \text{for} \quad 1 \leqslant i \leqslant j \leqslant M. \tag{6}$$

DMAS beamformer can be considered as a beaming algorithm with a correlation process. In other words, the output of this beamformer is the spatial coherence of detected PA signals, and it is a non-linear beamforming

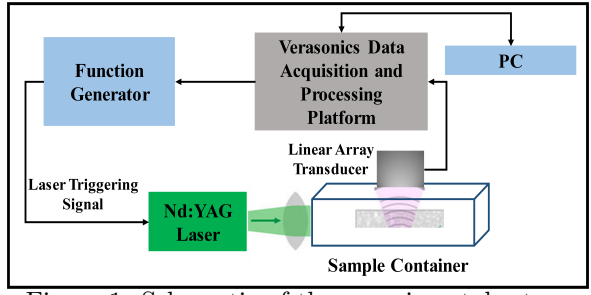
Figure 1: Schematic of the experimental setup.

algorithm. The expansion of DMAS algorithm can be written as follows:

$$y_{DMAS}(k) = \sum_{i=1}^{M-1} \sum_{j=i+1}^{M} x_{id}(k)x_{jd}(k) = x_{1d}(k)\underbrace{\left[x_{2d}(k) + x_{3d}(k) + x_{4d}(k) + ... + x_{Md}(k)\right]}_{\text{first term}}$$

$$+ x_{2d}(k)\underbrace{\left[x_{3d}(k) + x_{4d}(k) + ... + x_{Md}(k)\right]}_{\text{second term}} + ... + x_{(M-2)d}(k)\underbrace{\left[x_{(M-1)d}(k) + x_{Md}(k)\right]}_{\text{(M-2)th term}} + \underbrace{\left[x_{(M-1)d}(k).x_{Md}(k)\right]}_{\text{(M-1)th term}}.$$

(7)

where $x_{id}(k)$ and $x_{jd}(k)$ are delayed detected signals for element $i$ and $j$, respectively. As can be seen, there is a DAS in every terms of the expansion, and it can be used to modify the DMAS beamformer. In every terms of (7), there exists a summation procedure which is a type of DAS algorithm. It is proposed to use EIBMV adaptive beamformer for each term instead of DAS. In other words, since DAS is a non-adaptive beamformer and considers all the calculated samples for each element of the array the same as each other, consequently, the acquired image by each term is a low quality image with high levels of sidelobe and broad mainlobe. In order to use EIBMV instead of each DAS in the expansion, we need to carry out some modifications and prepare the expansion in (7) to adopt the EIBMV. The necessary modifications are extensively explained in reference[21] and are out of the scope of this paper.

The output of EIBMV-DMAS can be written as follows:

$$y_{EIBMV-DMAS}(k) = \sum_{i=1}^{M} w_{i,new} \underbrace{\left(x_{id}(k)\left(\sum_{j=1}^{M} w_j(k)x_{jd}(k)\right) - w_i(k)x_{id}^2(k)\right)}_{i_{th} term},$$

(8)

where $w_{i,new}$ is the calculated weight for each term in (8), $w_i$ is the calculated weights for each term in (7) while the steering vector in the formula of EIBMV is a vector of ones. It should be noticed that when there is a multiplication, resulting in the squared dimension, the method mentioned in (3) and (4) is used to prevent the squared dimension

## 3. EXPERIMENTAL RESULTS

A linear-array of PAI system was used to detect the PA waves, and the major components of system include an ultrasound data acquisition system, Vantage 128 Verasonics (Verasonics, Inc., Redmond, WA), a Q- switched Nd:YAG laser (EverGreen Laser, Double-pulse Nd: YAG system) with a pulse repetition rate of 25 $Hz$, wavelength 532 $nm$ and a pulse width of 10 $ns$. A transducer array (L7-4, Philips Healthcare) with 128 elements and 5.2 $MHz$ central frequency was used as a receiver. A function generator is used to synchronize all operations (i.e., laser firings and PA signal recording). The data sampling rate was 20.8320 $MHz$. The schematic of the designed system is presented in Figure 1. A wire has been used as the phantom of imaging where the surface of the array was perpendicular to the imaging phantom. Thus, it is expected to see a cross section of the wire, which would be like a point. The reconstructed images using the concerned beamformers are shown in Figure 2 where DAS leads to a low quality image having a low resolution along with high sidelobes. Moreover, the effects of the noise of the imaging system and artifacts affect the reconstructed image. DMAS results in a higher noise and artifacts suppression and improves the image quality as shown in Figure 2(b). However, the resolution is not good enough in comparison with EIBMV, which can be perceived by comparing the images shown in Figure 2(b) and Figure 2(c). Using EIBMV-DMAS results in a higher resolution, and lower artifacts and sidelobes compared to other concerned beamformers, as shown in Figure 2(d). To have a more precise evaluation, the lateral variations of the images shown in Figure 2, are shown in Figure 3(a) along with a zoomed version (Figure 3(b)). The level of sidelobes for DAS, DMAS, EIBMV and EIBMV-DMAS are about -38 $dB$, -55 $dB$, -127 $dB$ and -177 $dB$, respectively. To put it more simply, EIBMV-DMAS outperform DAS, DMAS and EIBMV in the term of sidelobes about 139 $dB$, 122 $dB$ and 50 $dB$, respectively. The narrower width of the mainlobe, resulting from EIBMV-DMAS compared to other beamformers, can be clearly seen in Figure 3(b). Although the resolution enhancement of EIBMV-DMAS is not significant compared to EIBMV, the sidelobes degradation is considerable. To have a quantitative comparison, the signal-to-noise ratio (SNR) has been calculated using the formula explained in reference.[17] The calculated SNRs can be seen in Table 1. As shown, the proposed method results in a higher SNR compared to other beamformers where it outperforms DAS, DMAS and EIBMV of about 65 $dB$, 41 $dB$ and 18 $dB$, respectively. All the experimental results show the superiority of the proposed method.

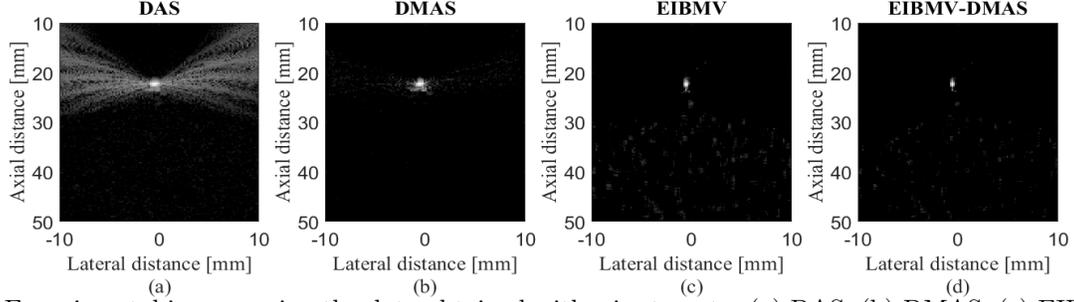

Figure 2: Experimental images using the data obtained with wire targets. (a) DAS, (b) DMAS, (c) EIBMV, (d) EIBMV-DMAS. All images are shown with a dynamic range of 60 $dB$.

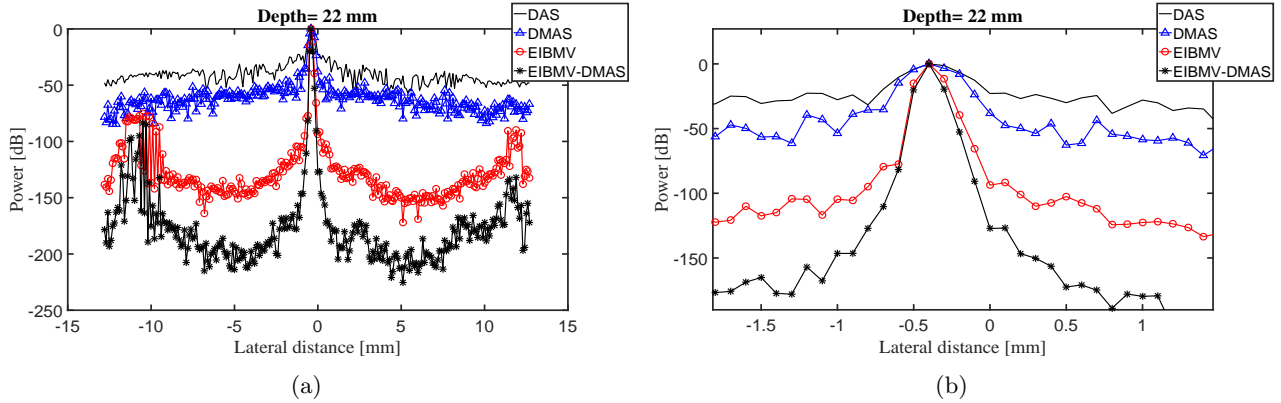

Figure 3: (a) Lateral variations of DAS, DMAS, EIBMV and EIBMV-DMAS at the depth of 22 $mm$ of the images shown in Figure 2. (b) A zoomed version of (a) for a better comparison.

## 4. CONCLUSION

In this paper, the combination of DMAS beamformer and EIBMV adaptive beamforming algorithm has been experimentally examined. The proposed method was quantitatively and qualitatively evaluated, and it was shown that it outperforms DMAS and EIBMV in the terms of sidelobes, width of mainlobe and SNR. EIBMV-DMAS improved the sidelobes, in comparison with DAS, DMAS and EIBMV, for about 139 $dB$, 122 $dB$ and 50 $dB$, respectively. Moreover, SNR was improved by the proposed method of about 158 %, 63 % and 20 %, respectively.

Table 1: SNR ($dB$) values at the depth of 22 $mm$.

| Depth($mm$) | DAS | DMAS | EIBMV | EIBMV-DMAS |
|---|---|---|---|---|
| 22 | 41.76 | 65.24 | 88.88 | 106.96 |